# Neural Networks Assist Crowd Predictions in Discerning the Veracity of Emotional Expressions

Zhenyue Qin, Tom Gedeon and Sabrina Caldwell

The Australian National University, Canberra, Australia
{zhenyue.qin, tom.gedeon, sabrina.caldwell}@anu.edu.com

**Abstract.** Crowd predictions have demonstrated powerful performance in predicting future events. We aim to understand crowd prediction efficacy in ascertaining the veracity of human emotional expressions. We discover that collective discernment can increase the accuracy of detecting emotion veracity from 63%, which is the average individual performance, to 80%. Constraining data to best-performers can further increase the result up to 92%. Neural networks can achieve an accuracy to 99.69% by aggregating participants' answers. That is, assigning positive and negative weights to high and low human predictors, respectively. Furthermore, neural networks that are trained with one emotion data can also produce high accuracies on discerning the veracity of other emotion types: our crowdsourced transfer of emotion learning is novel. We find that our neural networks do not require a large number of participants, particularly, 30 randomly selected, to achieve high accuracy predictions, better than any individual participant. Our proposed method of assembling peoples' predictions with neural networks can provide insights for applications such as fake news prevention and lie detection.

## 1   Introduction

Acted emotions are facial expressions whose performers do not carry genuine feeling [1]. By using acted emotions, human beings attempt to convince others that they are experiencing the pretended mental state. For example, sales staff act smiles to their customers in order to express friendly attitudes. Commonly acted emotions in our daily life include anger, surprise, fear, and happiness. These four emotions are assumed to be recognizable regardless of cultural background [2]. The ability to distinguish between acted and genuine emotion expression can aid in lie detection, advertising effect assessment, and other applications [3, 4].

Research indicates that in general, human beings perform poorly in verbally differentiating between genuine and acted emotion expressions [1]. Hossain et al. reported that the accuracy of participants' verbal responses in respect of smile veracity was approximately 60% [1], which was better than random guessing (50%). However, their work did not investigate the accuracy of a collection of participants' responses to a particular emotion expression as a whole. Instead, they considered only the accuracy of an individual's reply. That is, an individual might demonstrate poor performance in discerning the emotion veracity. Nevertheless, utilizing a majority response from a group of individuals as their final collaborative answer may present a higher-accuracy response.

Moreover, researchers in social sciences have shown the promise of utilizing the crowd to predict future events [5]. Their work indicated that the prediction results from crowds, of five US presidential elections were more accurate than the traditional pools 74% of the time [6, 7] . Due to this higher accuracy of crowd forecasting, it has been adopted by a range of industries including healthcare companies, technology corporations and so on [8].

However, the previous applications of crowd forecasting weighted all the participants' responses with the same importance. That is, they did not acknowledge that people's predicting capabilities can vary. To supplement this defect, in 2015, researchers improved the traditional forecasting methodology by extracting top-performing predictors through an array of prediction tasks and assigning them into elite teams called superforecasters [9]. These teams composed of the selected top-performers demonstrated a 50% greater accuracy than traditionally assembled crowd forecasting teams [10].

Nonetheless, all of these previously conducted experiments on crowd forecasting provided financial rewards for correct predictions. As such, a natural question to ask is whether crowd forecasting can still give high-quality predictions without any payments. A study conducted in 2004 showed a positive result by comparing the prediction qualities from providing participants with real and play money [11]. That is, recompense is not essential for stimulating high-quality predictions. Furthermore, both crowd predictions exhibited more significant power than individual humans [11, 12].

Inspired by the success of crowd predictions, we would like to know whether aggregating a group of people's responses will lead to better answers for discerning emotion veracity. Furthermore, considering Meller et al.'s results on the success of utilizing elite teams [9], we hypothesize that some participants may also demonstrate better capability in discerning pretended emotion expressions than others. We would also like to know whether a team of elites detecting emotion veracity will present higher accuracy than a team consisting of average participants.

Moreover, it is also worthwhile investigating if a neural network can distinguish better emotion discerners and assign higher weights to their responses. In contrast, there may also exist people who are particularly poor in the emotion recognizing task and who tend to always give incorrect answers. We wish to know whether neural networks can learn to assign them negative weights to flip their responses so that even these poor performers can make contributions to a higher accuracy.

Furthermore, we would also like to know the minimum number of participants to achieve a highly accurate result, which can make contributions to reducing the data collection cost when utilizing this technique. Additionally, we also want to know whether elites who can discern one emotion expression accurately can still give a high-accuracy performance for detecting other emotions. Similarly, we wish to study the transfer learning ability [12] of our neural networks. That is, whether a neural network built for discerning one emotion can also work well on other emotions. In the future, we wish to discover the potential similarity between distinguishing genuine and acted emotion expressions and identifying fake news. Ideally, the methodologies in discerning emotion veracity can be applied to other forms of veracity detection and hence be useful in the prevention of fake news spread.

## 2      Methods

### 2.1     Stimuli

Previous work conducted by Hossain and Gedeon investigated people's ability to distinguish acted and genuine smiles, using both verbalized descriptions and pupil dilation responses from the participants [13]. Chen et al. extended this research to investigate the emotion of anger [1] with the same methodology. However, Hossain and Gedeon used minimal stimuli [13], where videos are cropped to only include the faces [14], while Avezier et al. suggested that the contextual backgrounds for displaying emotional expressions are essential to differentiating emotions [15]. Therefore, both this paper and the research done by Chen et al. include some contextual backgrounds and the stimuli they provide.

The raw videos for the experiments were sourced from YouTube. Each emotion type consisted of 20 videos, which were selected considering balancing ethnicity, gender, and background context to reduce unnecessary noise in the experiment. Genuine emotion expressions were collected from reality television shows and acted ones were obtained from movies containing similar scenes.

### 2.2     Neural Networks

A simple feedforward neural network [16] is utilized to give participants different weights in order to value high/low-quality responses differently. Specifically, the neural network contains 117 input neurons, corresponding to 117 participants' answers to a stimulus video. In addition, we also provide the neural network with the correct label of that video. That is, whether the video is a genuine or acted emotional expression, 1/0 respectively. The number of inputs corresponds to the number of human predictors. For example, if there are 117 participants, then the input layer will have 117 nodes. Each input corresponds to a participant's prediction. Specifically, 1 for genuine and 0 for acted emotions. This structure guarantees that the network can learn from the behaviour of specific individuals, as each input represents one person.

The network has a single hidden layer consisting of 10 neurons, which use the logistic activation function, and achieved excellent results. We attempted various number of hidden neurons, up to 20 which all produced similar results. The network classifies its inputs, corresponding to the likelihood of the emotion presented in the video being genuine or acted emotion. We train our neural network with stochastic gradient descent [17] and cross entropy [18]. We trained for 5000 epochs based on an initial pilot, with a learning rate is 0.01. Additionally, we utilize leave-one-video-out cross-validation to test the performance of our neural networks. That is, we train a neural network to learn the reliabilities of each participant using the first 19 videos, and test with the last one video, and repeat 20 times, reporting average results. Thus, we can be confident that our networks are not overfitting. We note that we used results for 80 videos in total, being 4 emotions x 20 videos per emotion. This is a considerable dataset for human data: 117 people x 80 videos x 2 mins (to watch a short video and decide on veracity, occasional short rests) = 312 hours.

### 2.3    Elite Detectors

Similar to superforecasters who are teamed up to give accurate predictions on future events [9], we also selected elites from the participants who demonstrated higher individual accuracy in discerning emotional veracity. Afterward, we utilized the majority response from those elites as the unified decision on a particular emotional expression. We also tested a range of elite sizes in order to find the best-performing elite ratio.

## 3    Results

### 3.1    A collective voting approach can increase the accuracy of human ability to discern emotion veracity

Previous research indicates that the individual capability of discerning emotion veracity, including smile and anger, is little above randomly guessing [1, 13]. It is also expected that a similar level of accuracy will be obtained for fear and happiness [19, 20]. Although individuals tend to give low-accuracy answers for emotion veracity, our experiments revealed that by adopting the most common response among a crowd, the accuracy can be increased to 80% for determining the veracity of an emotional expression.

### 3.2    Teaming up elites can increase the accuracy of discerning emotion veracity

Preliminary analysis on participants' responses indicated that people's capability to discern emotion veracity varied. That is, there exist some people who demonstrated higher accuracies in detecting veracity of emotional expressions. Our results revealed that teaming up better emotion veracity detectors and averaging their responses can produce higher accuracies than considering all the participants' responses. We also found that an elite ratio of 5% would lead to the best performance in detecting emotional veracity, at an accuracy of approximately 94%, as Figure 1 indicates.

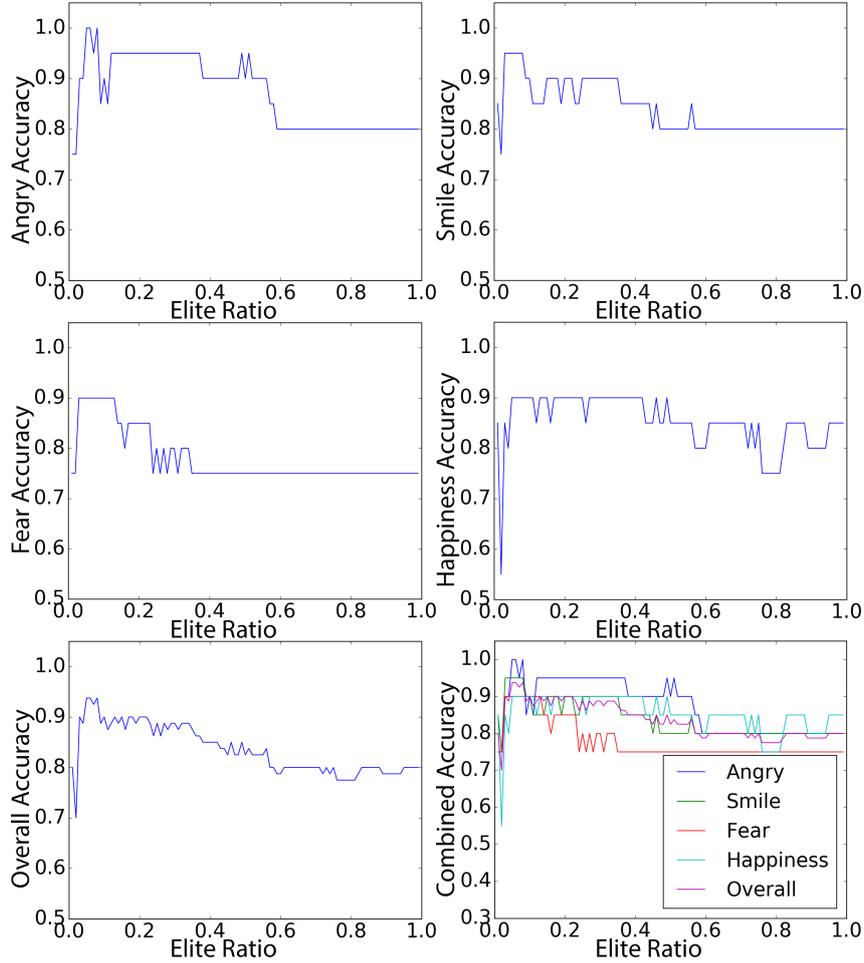

**Fig. 1.** Accuracy vs elite ratios

### 3.3 Neural networks assist in aggregating participants' responses to present higher accuracy

The neural network, specified in Section II-B, demonstrated high-level performance in discerning emotional veracity from participants' responses. Specifically, an overall 99.69% accuracy for all of four emotion types, tested by the leave-one-out cross-validation approach. The quantity of training data matters to the performance of neural networks, such that more training data can result in higher accuracies. This is shown in our results: increasing the number of folds in cross-validation leads to corresponding performance growth, as Figure 2 indicates. The shapes of the curves also tell us that, e.g, anger is most strongly recognized, and that of smiles, the least strongly. This is consonant with the Psychological literature of emotion recognition, see our Introduction, and a validation of our techniques that this just 'falls out of the data'.

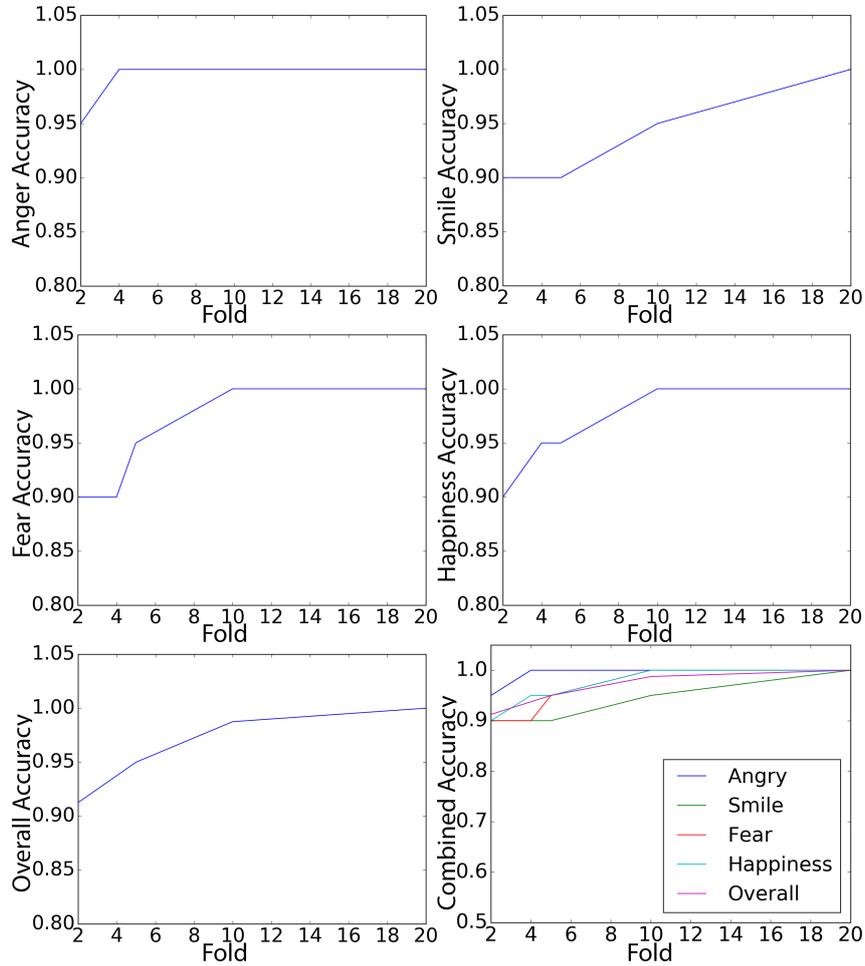

**Fig. 2.** Fold efforts

### 3.4 Neural networks that are trained from one emotion perform well on other emotions

In order to investigate the transfer learning ability of our neural networks, we trained them with one emotional expression data and tested their performance with the other three kinds of emotional data. For example, we trained a neural network with the anger data and tested using the smile, fear, and happiness data. As Table 1 indicates, our neural networks demonstrated high accuracy (90% on average) when identifying veracities of other emotional expressions. This suggests that emotions may share similarities in the way they are acted or that human have generalized veracity detectors.

Table 1. Train one emotion and test on the other three

| ANGER | SMILE | FEAR | HAPPINESS |
|---|---|---|---|
| Training | 0.75 | 0.9 | 0.9 |
| 0.95 | Training | 0.9 | 0.9 |
| 0.95 | 0.85 | Training | 0.9 |
| 0.95 | 0.95 | 0.9 | Training |

### 3.5 Combining all the emotion data to train neural networks still produces high-accuracy performance

In order to further investigate the similarities among predictability of veracity of the different emotional expressions, instead of dividing the data into four parts, corresponding to the four emotional expressions, and do the training as well as testing for each of them separately, we combined the data for all the four emotions. Sticking to the leave-one-out cross-validation approach, the accuracy increased to 100% with 20 reputational runs. Again, this implies that distinct emotions may share some likeness. It appears that people who are expert in distinguishing the veracity of one emotion are also excellent in discerning the veracity of other emotions. This can also be seen in Table 2, which shows the results of utilizing elites of one emotion to discern the veracities of other emotions.

Table 2. Elites of one emotion to predict others

| ANGER | SMILE | FEAR | HAPPINESS |
|---|---|---|---|
| Training | 0.8 | 0.9 | 0.65 |
| 1.0 | Training | 0.9 | 0.85 |
| 1.0 | 1.0 | Training | 0.9 |
| 0.8 | 0.85 | 0.8 | Training |

Figure 3 demonstrates the overlapping of the top n elites for distinguishing the four emotional expressions. For example, we separately picked the top 5 elites in discerning each emotion type. Among these 20 people (5 elites times 4 emotions), 20% of them were the same.

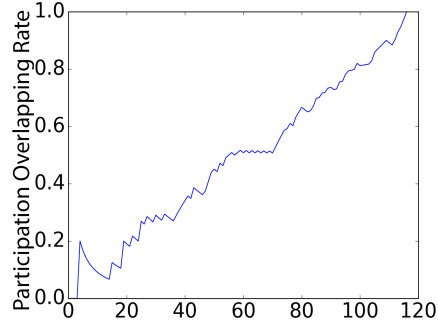

**Fig. 3.** Overlapping rate of elites for different participant numbers

### 3.6   The effects of participant numbers on accuracy

We wish to discover the relationship between accuracy levels and the numbers of participants. Finding a minimum number will minimize resource costs of using such crowd prediction techniques. To investigate, we randomly picked various numbers of participants from all the participants and repeated this process 20 times for each size of group.

For example, we randomly chose 40 participants from all 117 and tested the accuracy using the 40 responses of those selected participants. In order to reduce noise resulting from the random selection, we repeated the action of randomly picking 40 participants and testing their accuracies 20 times, and averaged the 20 accuracy results as the final accuracy for this size of group.

As Figure 4 indicates, when the number of participants increased, the accuracy also grew. Moreover, the growth became smoother when the total number of participants reached approximately 20. This suggests that a high accuracy on discerning emotion veracity did not require a large number of total participants. In our case, 30 participants lead to 99% accuracy of our neural networks in discerning the veracity of all four emotions.

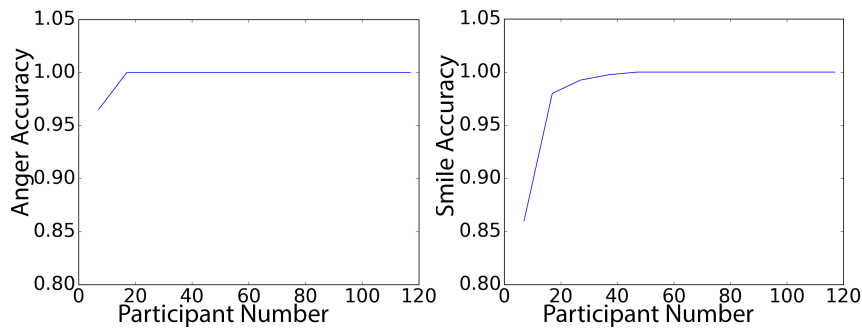

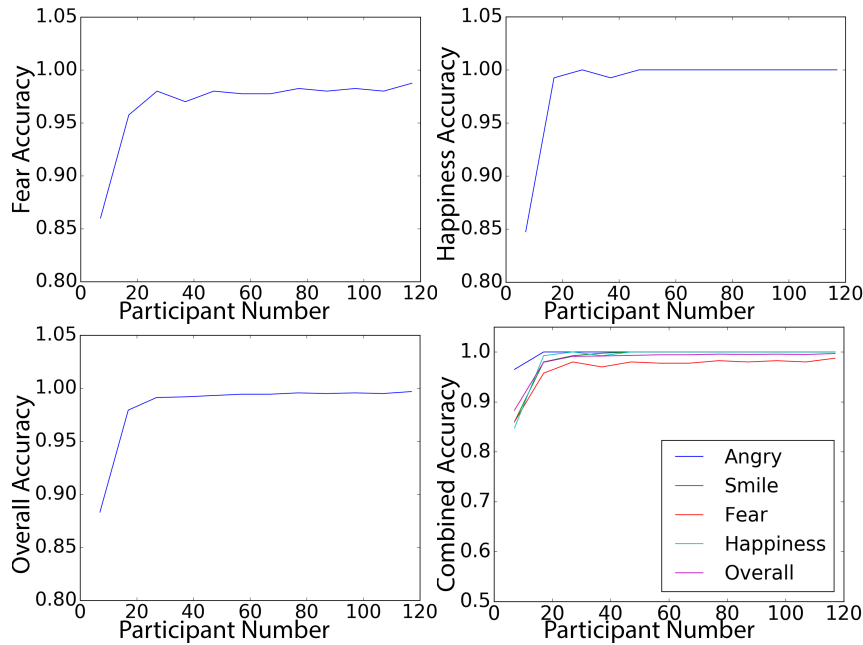

**Fig. 4.** Accuracies for different participant numbers

## 4 Discussion

### 4.1 An explanation on why neural networks can give high accuracy results in this problem

Our results indicated that neural networks can effectively aggregate various participants' responses and so identify the veracity of emotional expressions in order to give precise answers. We hypothesized that neural networks would assign positive weights to the responses of participants who give many correct answers and would assign negative weights to those participants who are more likely to give wrong responses. Thus, the wrong answers can still contribute to correct predictions. We suspected that this exploitation of incorrect responses explains why our neural networks performed better than teaming up elites.

In order to verify our hypothesis, we composed a dummy dataset in which the first three participants would always present correct answers, whereas the remaining seven would keep giving wrong answers.

After training with this dummy dataset, our neural networks would assign positive weights to the first three elements and negative weights to the remaining seven. Therefore, the hypothesis which states neural networks negated the responses of people who always present wrong answers in order to predict right results seemed to hold.

Furthermore, we wanted to know whether the weights assigned to each participant correlated to the person's individual discerning accuracy. For example, if a participant's accuracy in detecting emotional veracity is the highest, will the neural net-works also assign him or her the highest weight? Our further investigation could not defend a positive answer to this question. Specifically, with the neural networks trained using the dummy data above, although the individual accuracies of the first three participants were the same, their assigned weights varied significantly. In order to further understand this, we plotted the overlapping rate given the top *n* individuals in the real data. That is, among the top *n* participants sorted by their discerning accuracies and assigned weights, what is the proportion of overlapped individuals between these two? This is shown in Figure 5. Given approximately the top half of the individuals, i.e., the top 60 participants, the overlapping rate is 54%. In contrast, the probability of maintaining the same or higher overlapping rate through random selection was only 0.2610. This result was a little bit high to defend the null hypothesis, that it was very unlikely to randomly generate the same overlapping rate without neural networks' operation.

Overall, this suggests that neural networks do give positive weights to people who are apt to give true answers and negative weights to those who present false responses, however, it is not necessarily perfectly correlated in assigning to levels of accuracy.

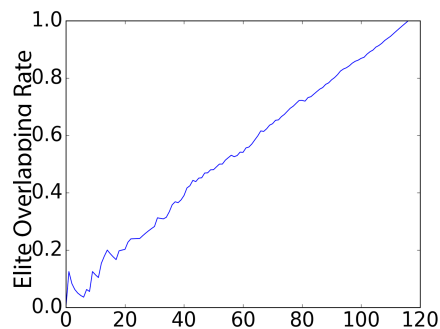

**Fig. 5.** Overlapping rate of individual assigned weights and accuracies

### 4.2   The difficulty of discerning the different kinds of emotion veracities vary

As Figures 2 and 4 indicate, anger converged faster than the other three emotional expressions. This implies that: 1. It does not require a lot of training data in order to discern the veracity of anger expressions; 2. It needs fewer participants in order to give accurate aggregated answers on distinguishing anger veracity. This means that it is easier to detect the veracity of anger expressions than others. This observation is consistent with the work of Mather and Knight [21], which indicated that anger is quicker to detect overall.

## 5   Future Work

In the future, we will extend our method to discerning the veracity of fake versus real news. First, we will conduct experiments on utilizing the wisdom of crowds collectively in order to recognize fake news. Second, if the crowd's accuracy on discerning fake news is satisfactory, privacy will be an issue to solve, as our neural networks need to be aware of the quality of specific participants' responses. However, people may resist the fact that they are found to often present incorrect answers or they may behave differently if they know that they are usually right.

## 6    Conclusion

In this paper, we aggregated peoples' verbalized responses to predict the veracity of emotional expressions comprising four universal emotions, namely anger, surprise, fear, and happiness. We discovered that by adopting collective voting instead of using individual responses, the accuracy of human discernment of emotion veracity overall could be increased from 63% to 80%. We also found that there exist people who demonstrate better abilities in ascertaining emotion veracity. By incorporating the responses from these 'elite' predictors, the overall accuracy could be further increased to 92% in collective voting. Finally, we introduced neural networks to aggregate participants' responses and obtained an overall 99.7% accuracy. Additionally, we found that training with one emotion data leads to high accuracies when testing with the other three kinds of emotion data using our neural network approach, a novel emotion transfer result. Our neural networks did not require a large number of participants for high-accuracy performance. A closer look revealed neural networks achieved these high-level results by assigning positive and negative weights to the participants who tended to give consistently good and bad answers, respectively. However, the weightings of participants in the neural networks may not simply reflect the ranking of participants' accuracies on ascertaining emotion veracity. In the future, we will utilize the same methods to investigate the feasibility of utilizing crowds to discern genuine and fake news.

## 7    Acknowledgments

The authors are grateful to Aaron Manson for access to the data, and particularly acknowledge his efforts in data collection.

12      Qin et al.